# Coupling agent based simulation with dynamic networks analysis to study the emergence of mutual knowledge as a percolation phenomenon

Julie Dugdale[1,2], Narjès Bellamine Ben Saoud[3,4], Fedia Zouai[3], Bernard Pavard[5]
(1) Université Grenoble Alpes; LIG, Grenoble, France
(2) University of Agder, Norway
(3) Laboratoire RIADI – ENSI Université de La Manouba, Tunisia
(4) Institut Supérieur d'Informatique, Université Tunis el Manar, Tunisia
(5) Université P. Sabatier, CNRS, IRIT, Toulouse, France

**Abstract**
The emergence of mutual knowledge is a major cognitive mechanism for the robustness of complex socio technical systems. It has been extensively studied from an ethnomethodological point of view and empirically reproduced by multi agent simulations. Whilst such simulations have been used to design real work settings the underlying theoretical grounding for the process is vague. The aim of this paper is to investigate whether the emergence of mutual knowledge (MK) in a group of co-located individuals can be explained as a percolation phenomenon. The followed methodology consists in coupling agent-based simulation with dynamic networks analysis to study information propagation phenomena: after using an agent-based simulation we generated and then analysed its traces as networks where agents met and exchanged knowledge. Deep analysis of the resulting networks clearly shows that the emergence of MK is comparable to a percolation process. We specifically focus on how changes at the microscopic level in our agent based simulator affect percolation and robustness. These results therefore provides theoretical basis for the analysis of social organizations.

1. **Background**

The aim of this paper is to propose a theoretical model for emergent organisations such as those often encountered in complex or degraded real-world cooperative systems.

Usually, organisations such as those in emergency control rooms, space control centres, and nuclear power plants are composed of a group of people interacting in a proximal space (real or virtual). In addition to verbal interactions, people also interact in an informal way through gestures and unobtrusive observations of the actions of others.

Such informal organisations are highly paradoxical; whilst there are many rules and procedures that constrain how the group should handle critical situations, the way that the activities unfold in dealing with an event is mainly non deterministic and unstructured. What is interesting is that this paradox is only apparent if we analyse the situation from a systemic point of view. There are many examples of this apparent paradox, for example in early field studies of Air Traffic Control (ATC) settings, Mell observed that even if verbal exchanges between air traffic controllers are fully constrained lexically and syntactically, in the real situation only 20% of exchanges follow the rules [18]. In a study concerning incident management in ATC, non-verbal communication and the use of informal artefacts were found to strongly structure coordination processes within the team [14].

The intuition behind these mechanisms is that mutual knowledge emerges more 'easily' with informal organisations than with normalised exchanges. This is largely due to the unobtrusive nature of an informal organisation where actors are gathering information as they need it and when their interlocutor seems available. Likewise, seemingly adhoc and informal broadcasting of information may also be selective and modulated by the context of the situation [5].

We are particularly interested in social cooperation and understanding collective behaviour where agents in a complex social system may rapidly share information. We have seen that mutual knowledge emerges when people are spatially close together and are willing to communicate. In such situations, information can propagate very rapidly with a minimal perturbation to on-going cognitive processes. Mutual knowledge has also been shown to contribute to the exceptional robustness of socio-technical systems [17] [19]. However there are many factors that can adversely affect emergent behaviours and are detrimental to cooperation. For example, if interactions between agents are impaired by noise, or if there are too many interlocutors, etc. the emergence of mutual knowledge may be drastically and rapidly reduced without actors being aware of the situation. Likewise, if people working in a group setting are too involved in their own individual activity, they are no longer able to overhear broadcasted communications and, as a consequence, group efficiency is drastically reduced (e.g. [17] [19]).

Despite some analytical studies that have examined information flows in organisational settings, there is no particular theory that can explain the advantages of such an informal organisation. Our own previous works in the design of such organisations, e.g. Air Traffic Control, emergency medical control room, and space control centres, used a multi agent simulation approach in order to empirically study the advantages and disadvantages of complex emergent organisations [24]. From these studies we were able to use these multi-agent models to design complex control rooms [23]. Whilst these works employed the general properties of complex systems theory, such as emergence and self-organisation, the underlying processes that led to advantageous proximal cooperation remained unclear. In this paper we suggest the percolation model as a good candidate for increasing our understanding of emergent processes in complex organisations.

Therefore the aim of this paper is:
- To confirm that the percolation mechanism is a relevant model for explaining the emergence of mutual knowledge in a group of people interacting locally in a non coordinated way,
- To understand what are the most important factors that affect the percolation of mutual knowledge (e.g. effect of population density, overhearing),

In section 2, we briefly introduce percolation as well as the two types of percolation processes (site and link percolation). A more in-depth discussion on mutual knowledge, as being generated as part of a complex socio-technical system, is given in section 3. Section 4 describes the methodological approach that couples both agent-based simulation and network analysis. The experiments performed and their results are given in section 5. Here we specifically focus on how changes at the microscopic level in our agent based simulator affect percolation and robustness.

Finally, section 6 draws some conclusions and discusses the implications of the work.

## 2. Percolation and its modelling process
### 2.1. Percolation and its types

The percolation mechanism was introduced by Broadbent and Hammersley and is a long studied phenomenon in the domains of physics and mathematics [3]. A classic example in physics is the study of porous material. Here the porosity of the material is modelled in terms of a probability that an open space exists between two sites of material. Above a certain probability threshold the material will be porous; conversely, below the threshold, it will be impermeable. Percolation is a non-linear process that can thus explain the emergence of connected clusters; clusters of empty space in the case of porous material. What is interesting in this concept is its non linear characteristic with singularities which explains how a process, above certain threshold, can drastically change its characteristics following universal rules. Furthermore these rules are independent of the domain of the process. Outside of mathematics, physics and materials science, the percolation mechanism has been observed in domains such as economics, ecology, biology, computer networks, epidemiology, and social science [6]. In social science and economics this concept has been used to better understand the interplay between local and global actors [30]. In the marketing domain, a social percolation model has been developed [28] where agents represent consumers situated in a social network with the aim of understanding potential markets for products.

The percolation problem considers a network in which each node is occupied with probability p and links are present only between occupied nodes. As the probability p increases, connected components, called clusters, emerge. Thus the percolation problem studies the properties of the clusters, and in particular their sizes, as a function of the occupation probability p. It is indeed intuitively clear that if p is small, only small clusters can be formed, while a large p will eventually engulf the structure of the original network. As a practical example to see the difference between site and link percolation, consider a network of people (agents) that communicate in the mountains through optical devices (such as a torch or laser). Here the agents will be represented as nodes and the links will be the optical signals. If the weather is perfectly clear and all agents are efficient, we will have a probability p of 100% of communication and the network will be very efficient in propagating the information. However, if the agents are totally efficient but with a low level of communication due to bad weather, the information will not propagate very far (we have link percolation). Alternatively, if the weather is perfectly fine, but the agents are inefficient (e.g. not very attentive) then the lack of efficiency in the network is related to the site's behaviour (not the links); we are then in a situation of site percolation. Finally, if both the agents and the weather are not efficient, we will have a situation of mixed percolation. From this simple example, we can better understand the difference between site and link percolation [27].

### 2.2. Modelling site and link percolation processes

Let us first consider what happens in site percolation for a two-dimensional lattice such as the one depicted in Figure 1. In Fig 1.a, the probability that a site is active is 0.4. We can see that some sites or nodes are isolated, whilst others are contiguous, forming clusters[1] yet there is no cluster that fills the entire network. In Fig 1b, we reach the critical probability (p=0.6) when the network is almost filled by connected sites (if a porous material were being modelled, we would say that we have reached the probability threshold that makes the material porous). In Fig 1.c, the entire network is connected.

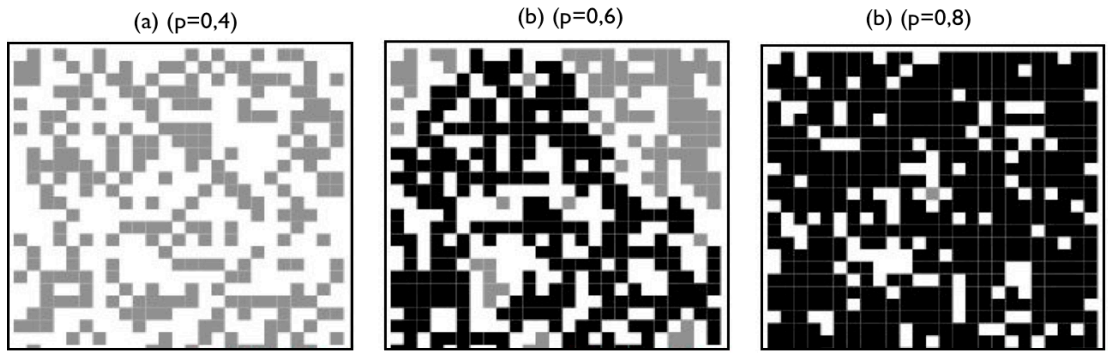

Figure 1: An example of site percolation for a square network. Each site is occupied, shown as black squares, with a probability p. Fig 1.a: when p is small (p=0.4), only little clusters of occupied sites are formed. Fig 1.b: At percolation probability (p= 0.6), a large cluster starts to emerge. Fig 1.c: Above the percolation threshold, the cluster invades the entire space. Note that if the dimension of the network is infinite, the cluster will also be infinite above the threshold [22]

Figure 2 shows the same phenomena but for a link network. In this type of network, only the links between sites are considered.

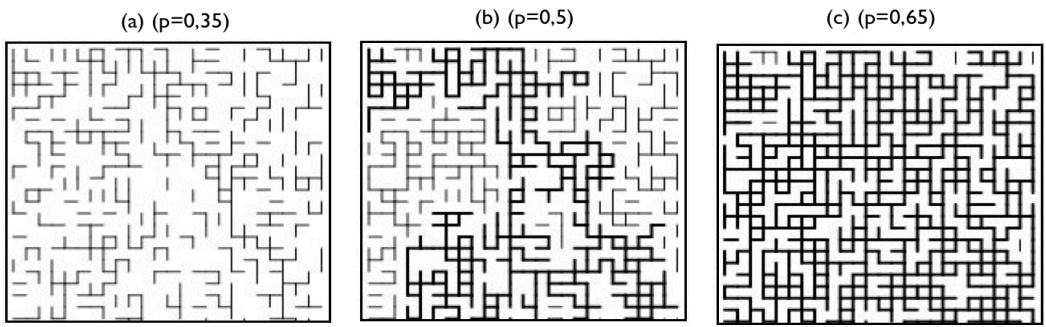

Figure 2: The same percolation phenomena in a link network [22]

To study percolation theoretically, we usually consider infinite networks that exhibit interesting properties such as:
- There is a critical p (denoted by $p_c$) below which the probability of having an infinite cluster is always 0 and above which the probability is always 1
- In networks of more than two dimensions, only simulation can approximate the percolation threshold $p_c$, it is not possible to calculate it
- The model has a singularity at the critical point $p = p_c$ believed to be of power-law type

If P(p) is the probability of percolation, we could say that:

---

[1] The concept of clustering refers to the tendency observed in many natural networks to form cliques in the neighborhood of any given link.

[2] We adopt the usual network terminology where an 'edge' refers to a link in a non-directed network (i.e. the meeting network) and where an 'arc' refers to a link in a directed network (i.e. the knowledge network). We will

P(p) =0 if p< $p_c$ (all clusters have a finite size)
P(p) >0 if p> $p_c$ (a giant cluster appears)
P(p) is a usually an increasing function with an exception point at the percolation threshold

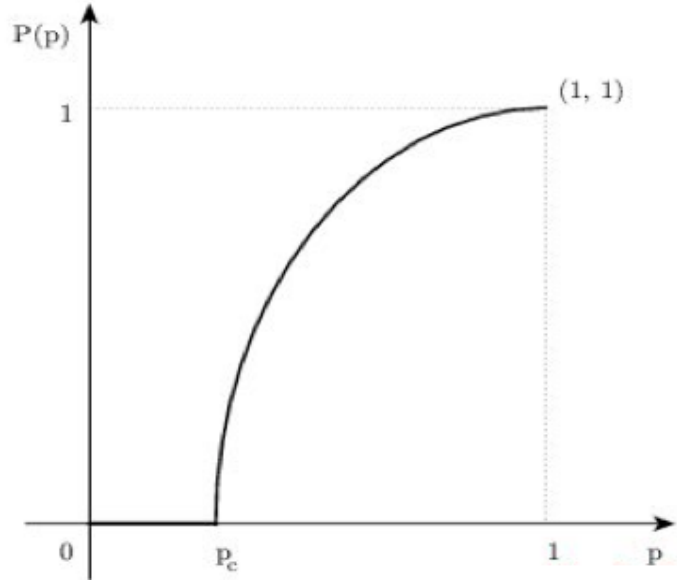

Figure 3: Probability P(p) for a node to belong to the infinite percolating cluster as a function of the occupation probability p. We can see a singularity at the percolation threshold $P_c$ [15].

Above the percolation point, the system exhibits invariant behaviour. The exact value of the critical exponents does not depend on the fine details of the percolation model. In general, they just depend on the system's dimensionality and symmetries of the order parameter. Thus, while the exact value of $P_c$ depends on the lattice geometry, the critical exponents do not. This universality also means that for the same dimension independent of the type of the lattice or type of percolation (e.g., link or site) the fractal dimension of the clusters at $P_c$ is the same. If we call $P_G$, the probability for a node to belong to the infinite percolating cluster, we have [2, p. 125]:

$$P_G \sim (p - p_c)^\beta$$

This means that $P_G$ follows a power law scaling from close to $P_c$. The scaling law expresses the insensibility of the characteristic quantities in a percolation process to the local and microscopic details around the critical value $P_c$.
The study of the percolation transition as a function of the connectivity properties of generalised random graphs finds a convenient formulation in the generating functions technique [7], [20].
Barrat & al. [2] report conditions for a giant cluster to arise in graphs that have a local tree structure with no cycles. They consider an uncorrelated network with degree distribution P(k) and compute the probability q that a randomly chosen edge leads to a vertex of degree k. This probability is written as the average over all possible degrees k of the products of two probabilities: (i) the probability that the randomly picked edge leads to a vertex of degree k and (ii) the probability that none of the remaining edges lead to a vertex connected to a giant cluster. This leads to the condition for the heterogeneity parameter (K) of the network [2] :

$$K = \frac{\langle k^2 \rangle}{\langle k \rangle} > 2$$

For a directed graph, the percolation phenomenon is also shown by the emergence of a large cluster in a network. Such a network is a giant cluster if:

$$K = \frac{\langle K_i . k_o \rangle}{\langle k_i \rangle} > 1$$

These are the criteria used to assess percolation in our generated dynamic networks.

## 3. Mutual knowledge and complex socio-technical systems

Mutual knowledge (MK) is knowledge that communicating parties share in common and that they know they share [16]. Mutual knowledge is also broadly referred to as 'common ground' and is an integral part in coordinating actions and collaborative decision-making [8, 9, 10].

Whilst MK is a very important concept it is also an ambiguous notion because it depends on the observer's capability and the richness of the media that people use to communicate. If we consider people interacting through email messages, the MK will depend, not only on the explicit messages exchanged, but also on the inferences that each agent may make when they receive the message.

In more interactive situations like face to face communications, MK will not only depend on verbal exchanges but also on non verbal ones such as gesture and posture. Physical constraints (such as the distance between agents) and the artefacts that mediate the communication can also affect MK.

It is important to understand that MK, like all emergent processes, needs an observer to identify it [1]. Thus the emergence of social structures in animal societies needs a human-being in order for it to be described. The actors themselves cannot see the global picture. In addition, agents are usually not conscious of the richness of their communication processes and an external observer is needed to assess the propagation of information as well as the structure of the information itself.

MK is therefore an abstract concept that may be far removed from real agents real representation. Nevertheless, in some ways it represents the knowledge that is accessible at a certain time (analogous to potential energy in physical systems).

Extensive analysis of work situations may lead to an acceptable understanding of the parameters that determine MK. If we consider a group of people interacting verbally in a room, we may first imagine that communication content, distance, noise, background knowledge or the goals of each agent are good candidates for MK parameters. In real situations the efficiency of proximal cooperation may depend on very subtle parameters such as how to direct information to a specific group of people without bothering others, or how to broadcast information dynamically to different groups of people without interfering with their on-going cognitive processes.

In real situations we can also observe that group efficiency behaves in some situations like a percolation process. Below a certain threshold of activity (or noise, workload, etc.), the group 'efficiency' is 'boosted' by its proximal local interaction. Above this threshold the group may experience a drastic loss of performance, usually without understanding the reason for it. In our previous works this situation was often observed in emergency control rooms. If the workload is acceptable, the room is not too noisy and overhearing process is efficient. We can then observe the 'percolation' phenomena: everybody is aware of everything and the network is efficient. However, when the workload increases, the noise increases and the overhearing capabilities are low. The same group of people may no longer be aware of other activities in the group and the efficiency of the group is low [4].

In order to understand better this process we previously developed empirically based simulators using a multi agent approach. The approaches varied from using a very detailed cognitive model of interaction [11] to more analytic approaches taking into account a limited number of parameters (such as the distance between actors, the radius of communication, the type of knowledge, and the rate of information exchange).

In one of these models, agents where moving randomly in their environment and each time they met they exchanged all their knowledge about the group to which they had initially belonged. MK was defined as the total amount of knowledge that was shared by each agent at each instant. We observed that the emergence of MK varied in a non-linear way with the size of overhearing capabilities (Fig. 4). However we were not able to formally identify this process as a percolation phenomenon. The aim of the next section is to explain our methodology in order to identify this process as a percolation process.

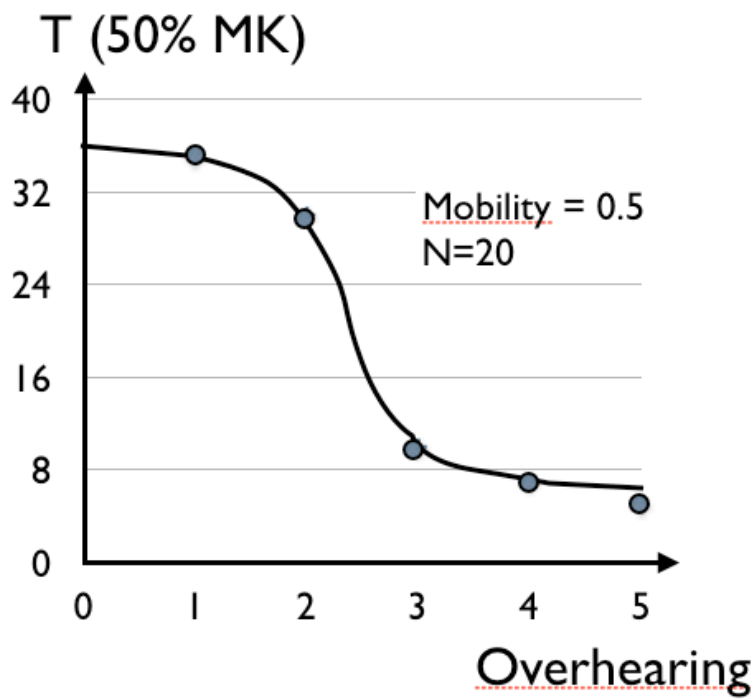

Figure 4: Evolution of MK (Time to reach 50% of Knowledge) in relation to the overhearing capability. M= Mobility of agents, N = number of agents

## 4. Methodological approach

### 4.1. Overview

The approach to the study of percolation was conducted in 4 phases, as shown in figure 5.

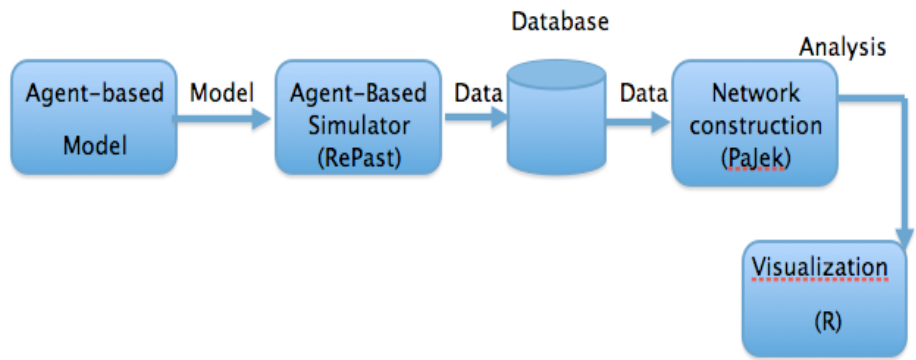

Figure 5: Workflow of our percolation study approach.

The agent-based model characterises the behaviour of the agents in their environment. The model was implemented in the Repast toolkit [26] and data from each time step was generated and stored in a database. Then, the data is used to construct a series of networks using Pajek [21], which is then analysed and visualized in R, which is a software package for statistical computation and graphics (R software [25]).

We generate a set of networks where each network represents one iteration of the simulation. The number of nodes in each network is equal to the number of agents in the simulation run, which is kept constant for each experiment.

The changes from iteration to iteration of the simulation are thus reflected in set of generated networks. Hence, we can see the dynamics of the network over time and specifically how the links between the nodes change. Graphs showing the evolution of the networks are generated using R [25].

**4.2. Agent-based model and simulator**

In this section we describe how the percolation mechanism has been modelled using an agent-based approach. Agent-based simulation is based on the idea that it is possible to represent computationally the behaviour of entities interacting in a world and that emergent phenomena can emerge as a result of these interactions. This approach therefore offers us a powerful tool in investigating macroscopic behaviours that result from interactions at the microscopic level. The model simulates the case where populations of agents, belonging to different groups, interact within an environment. When an agent meets another agent within its perception radius, they exchange information concerning their groups.

We are interested in measuring the level of MK over time. Intuitively this social phenomenon arises through the propagation of information within a population. The evolution of two artificial social networks (a Meeting network and an Exchanged Knowledge network) was studied to see if these networks exhibit percolation. As mentioned in section 1, the percolation phenomenon is characterised by the emergence of a large cluster in a network with the following property:

$$K = (<k^2>/2<k>) > 1$$
where k is the number of mean links between the nodes.

If a network exhibits this condition at time t, then it is in a percolation phase.
We also test how local factors impact this phenomenon at the macroscopic level. The factors taken into account are: the size of population; the individual properties of the agents (such as their ability to overhear, their propensity to forget information, and their mobility); environmental properties such as its dimension, and the nature of the exchanged information, such as the frequency of group changes are also considered.
The model consists of N agents (minimum $10^3$) moving randomly in an environment represented as a grid. Each cell of the grid is the same size and may contain one or several agents. Agents are initially randomly assigned membership of a specific group, which may change over time. Each time an agent meets another agent within its neighbourhood, it provides information concerning its own group as well as the group names of agents that it knows. An agent A can therefore know the agents that agent B knows without having met them. If there is conflicting information, e.g. agent A believes that agent C belongs to group 1, whereas agent B believes that it belongs to group 2, then the most recent information is used for updating. Each agent is characterized by the following information:

- ID: Each agent is assigned a number which uniquely identifies that agent
- Group membership: Each agent is randomly assigned to a group at the start of the simulation. An agent may change group randomly over time; a simulator parameter is considered for this and is set to false by default.
- List of agents met: each agent keeps a list of all of the agents that it has met. Each element of the list is composed of three fields: the ID of the agent that has been met, the group membership of the agent that has been met, the number of times that the two agents have met. Initially the list contains only one item (the ID and the group membership of the agent itself and the number 1 - signifying that the agent has met itself once). If we assume that we have 2 agents: A and B, each time A meets B there are two possible cases: A is meeting B for the first time, in this case A adds a new element to its list containing the ID and group of B, and the number 1. If A has already met B, it will simply increment the frequency.
- List of known agents: each agent keeps a list of all the agents that it knows either directly or indirectly. Again each element of the list is composed of three fields: the ID and group of the agent, and a timestamp indicting when this information was obtained. As with the list of agents met, this list initially contains information only about the agent itself. When the agent meets another agent, say agent B, it adds details of the agents that B knows to its own list and updates any incorrect information, such the change of group of an agent. Thus if there is an element in B's list that is not in A's list, it will be added to A's list (without changing the time). Otherwise, if the element is already in A's list, A checks to see if the group membership is the same in both lists and if it is, A uses the most recent time. If the group is different, A checks the time in both lists and uses the group corresponding to the most recent time.

In addition each agent has the following abilities:
- Move: agents move randomly in the grid according to a certain speed that was specified at the start of the simulation.

- Meet: Agent A meets Agent B if they are within each other's perception radius. This radius is the same for all agents and is specified at the start of the simulation. The radius represents the 'overhearing range' of an agent.
- Talk (to agents they meet): Agent A talks to Agent B means that A gives a list of all the agents and their groups that it knows to B. An agent can talk to several agents that it meets at the same time.
- Listen (to agents they meet): Agent A listens to Agent B means that A receives the list of all the agents and their groups from B.
- Forget: A forgets B means that A forgets what group B is a part of. Practically this corresponds to B being deleted from A's list of agents that it knows. The probability of forgetting is a variable set at the beginning of the simulation.
- Update incorrect information: since agents can change groups over time it is likely that an agent will, at some stage, have incorrect information about another agent's group. Therefore each time two agents meet, they compare lists and update incorrect information.

Table 1 summarizes the parameters and values that were used during the experiments

|  | Name | Abbreviation | Values |
|---|---|---|---|
| Variables | Dimension/density | d | $1000^2$, $5000^2$, $10000^2$ |
|  | Number of Groups | ng | 2 |
|  | Number of Agents | N | $10^3$, $10^4$, $10^5$ |
|  | Mobility | m | 100, 300, 600, 1000 |
|  | Overhearing | oh | 0, 1, 2, 3, 4 |
|  | Probability of forgetting | pforg | 0.001, 0.1, 0.5, 0.7 |
|  | Change group/changing information | chgr | True, False |
| Evolution functions | Speed of diffusion (knowledge network) |  |  |
|  | Meeting frequency (meeting network) |  |  |

Table 1: Parameters and values of the agent-based percolation simulator

Figure 6 below shows the simulator interface.

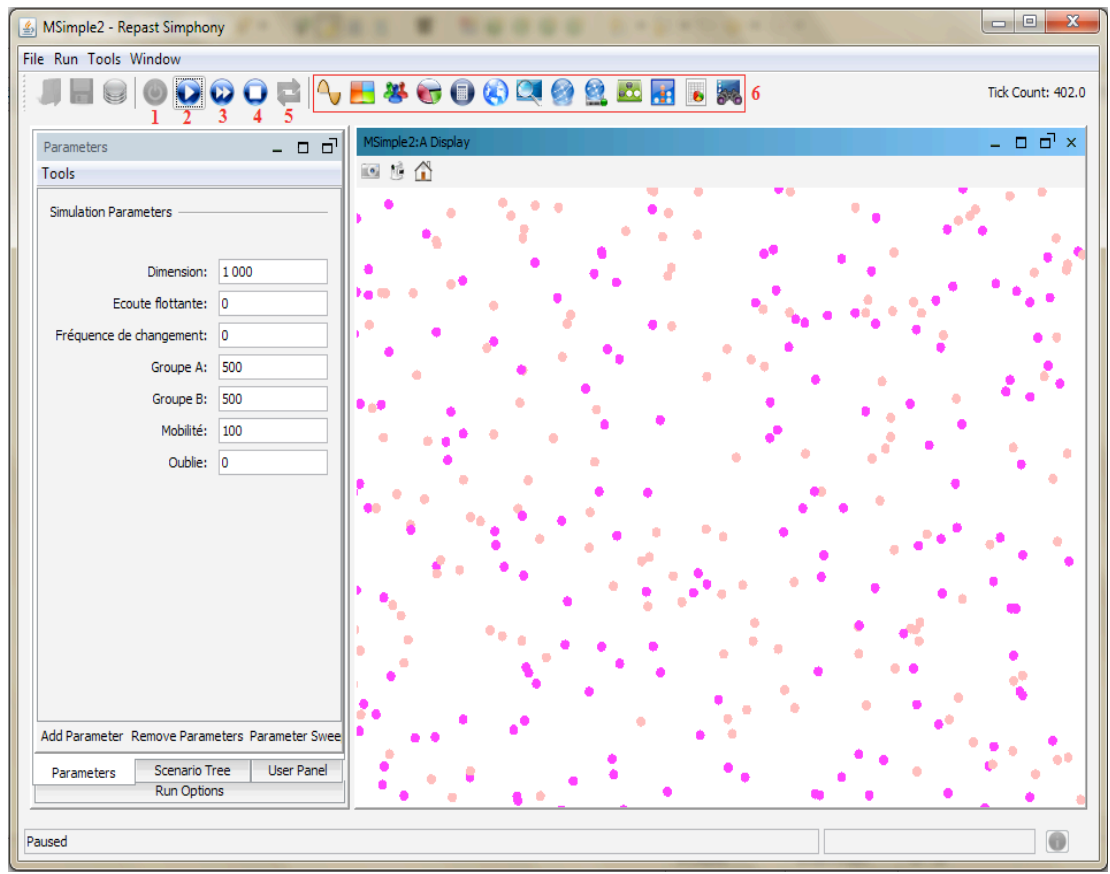

Figure 6: Agent based simulator interface at time t. The tab 'Parameters' in the bottom-left of the figure allows the user to enter the values for the parameters. The icons at the top of the figure allow the user to drive the simulation (e.g. initialize the model, start or stop the simulator or see a step-by-step iteration of the simulation, etc.). Once the simulation is complete, other buttons allow the user to send the results to external software packages, such as Pajek, R or even Excel to analyse the structural properties of the two networks generated by the simulation.

**4.3 Networks**

The interaction between agents in the environment is represented by two networks: a meeting network and an exchanged knowledge network. The meeting network is a weighted non-oriented graph where the nodes represent the agents. An edge between two agents with a weight x indicates that one agent has met the other x times since the start of the simulation. Conversely, the knowledge network is a directed graph where nodes represent the agents and where the arc[2] from agent $A_i$ to $A_j$ indicates that $A_i$ knows $A_j$ in the sense that $A_i$ has received, through previously meeting $A_j$ at time t, the group membership information about $A_j$. Note that because it is a directed graph, whilst $A_i$ may know $A_j$, $A_j$ does not necessarily know $A_i$.

---

[2] We adopt the usual network terminology where an 'edge' refers to a link in a non-directed network (i.e. the meeting network) and where an 'arc' refers to a link in a directed network (i.e. the knowledge network). We will use the term 'link' when we are referring to both networks.

We are interested in studying the impact of local factors and the structure of the network on the percolation phenomenon. At the beginning of the simulation we set the parameter values, e.g. the size of the population, the overhearing range of the agents, etc. At the end of each simulation we obtain two sets of networks: a set showing how the meeting network has evolved over time, and another one, showing how the knowledge network has evolved.

## 5. Experiments and Results

### 5.1 Knowledge network: analysing the percolation

The first objective is to look for a percolation process in knowledge network where the arcs between the sites indicate that agent $A_i$ knows agent $A_j$ (directed graph). The percolation phenomenon is shown by the emergence of a **large cluster** in a network.

Such a network is a giant cluster if:

$$K = \frac{<K_i.k_o>}{<k_i>} > 1 \quad \text{(Condition 1)}$$

If $K_i$ = input arcs (number of agents that know $A_i$) and $K_o$ = output arcs (number of agents that $A_j$ knows)

with the percolation condition :

$$P_sG = \frac{\text{Number of sites in the giant cluster}}{\text{Total number of sites in the network}} = f(P_s)$$

$$P_{sG}(p) \begin{cases} = 0 & \text{si } p < p_c \\ \approx (p - p_c) & \text{si } p > p_c \end{cases} \quad \text{(Condition 2)}$$

$P_sG(p)$ = the probability that a site belongs to the infinite cluster of the network.

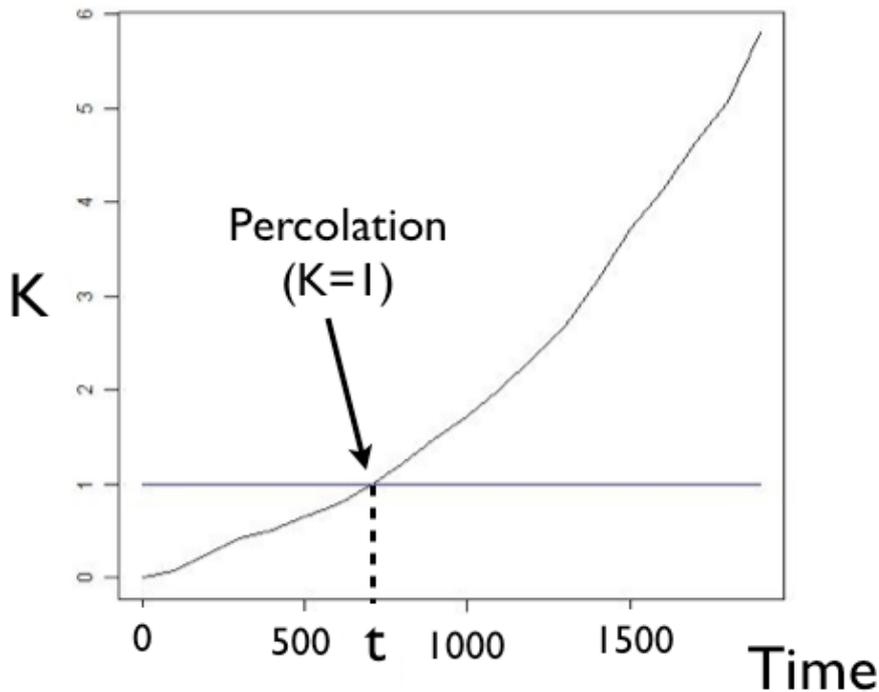

Figure 7: Evolution of the K parameter over time for a knowledge network of size $10^3$

If the network verifies this condition at time t, we will say that it is in a percolation phase. Figure 7 shows the evolution of K over time with a knowledge network of size $10^3$. The figure shows that K > 1 from t =700.

We will now use condition 2 to compute if our site network shows percolation characteristics. The expectation is that above a certain value of $P_s$ (probability that one node is connected i.e. it is linked to at least one other node) most of the sites will be connected and a giant cluster will emerge; the network will percolate.

In order to visualize site percolation, we draw $P_sG$, which is the probability that a site belongs to the infinite cluster of the network, in relation to $P_s$. Our aim is to find the point t that shows the percolation threshold (Fig. 8, A).
In figure 8, the values of $P_sG$ are largely greater than zero below the percolation threshold (shown by the red line) and the meeting network does not exhibit site percolation phenomenon. Concerning the knowledge network, the values of $P_bG$ are close to zero below the percolation threshold, showing that arc percolation exists.
As we can see, in knowledge networks, the percolation condition is not reached for site percolation (K>1 but $P_s$ >0). Conversely we can observe arc percolation (K=1, $P_bG \approx 0$). Figure 8, B shows that $P_bG$ begins to be positive at Pb=0.0006 (critical point). Thus, in the knowledge network if arc percolation is considered, $P_bG$ (i.e. the probability that an arc between agents belongs to a giant arc network) meets both the first and second conditions (Fig. 8).

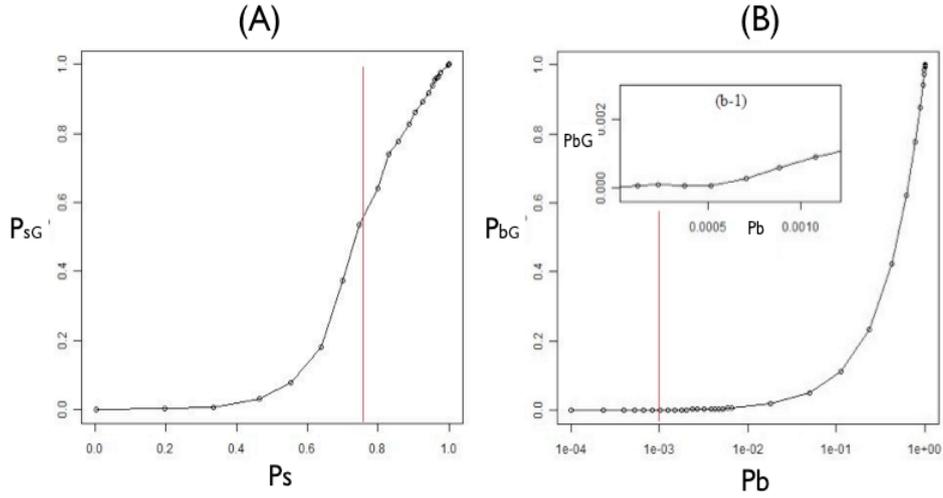

figure 8 (A): Probability that a site belongs to the infinite cluster of a network ($P_sG$) in relation to $P_s$ (probability that a site is active all over the network). The red line shows $P_s$ when K=1 (the condition for the existence of a cluster that percolates). (B): Probability that an arc belongs to the infinite cluster of a network ($P_bG$) in relation to $P_b$ (probability that an arc is active all over the network). The insert in figure 8, B, shows a magnified view at the percolation threshold. The size of the network = $10^3$.

Thus knowledge networks exhibit a arc percolation but no site percolation. In the next section we perform the same analysis with meeting networks.

**5.2 Meeting networks: Analysing the percolation phenomenon**

The same reasoning can be applied to meeting networks and the calculation of percolation condition becomes:

$$K \approx \frac{<k^2>}{2<k>} > 1$$

Figure 9 shows the graph Pbs = f(Ps) for a meeting network of $10^3$ nodes. Following the same steps, we can see that the meeting network exhibits a mixed percolation phenomenon (node and edge percolation). This result holds for all of the studied meeting networks (independent of initial variables).

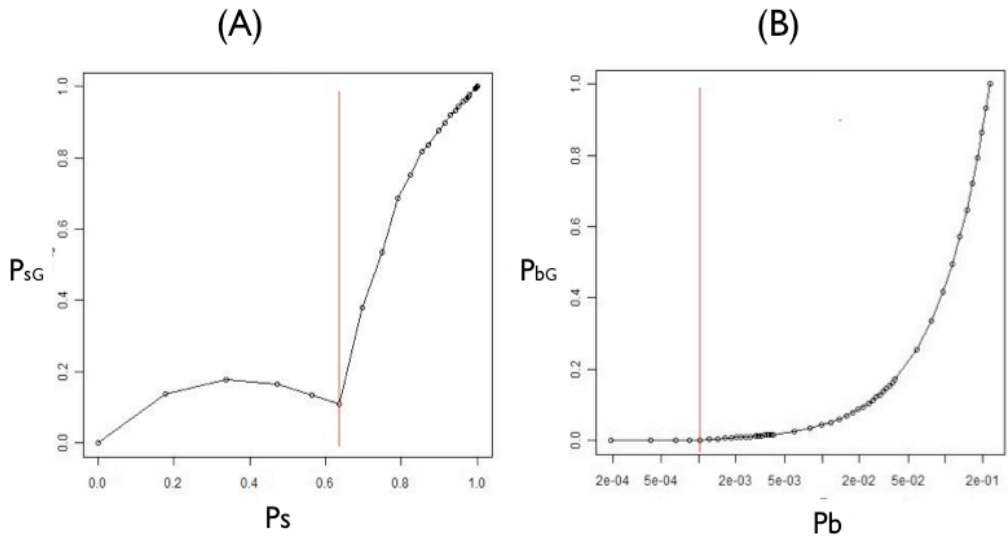

Figure 9 (A): Probability that a site belongs to the infinite cluster of the meeting network ($P_sG$) in relation to $P_s$ (probability that a site is active). (B): Probability that an edge belongs to the infinite cluster of meeting network (PbG) in relationship to Pb (probability that an edge is active). The red line shows Ps for where K=1 which is the condition for the existence of a cluster that percolates. Size of the network = $10^3$.

**5.3 Comparison between the knowledge network and the meeting network**
In order to see the similarities and differences between the knowledge network and the meeting network we can compare the respective graphs shown in figure 10.

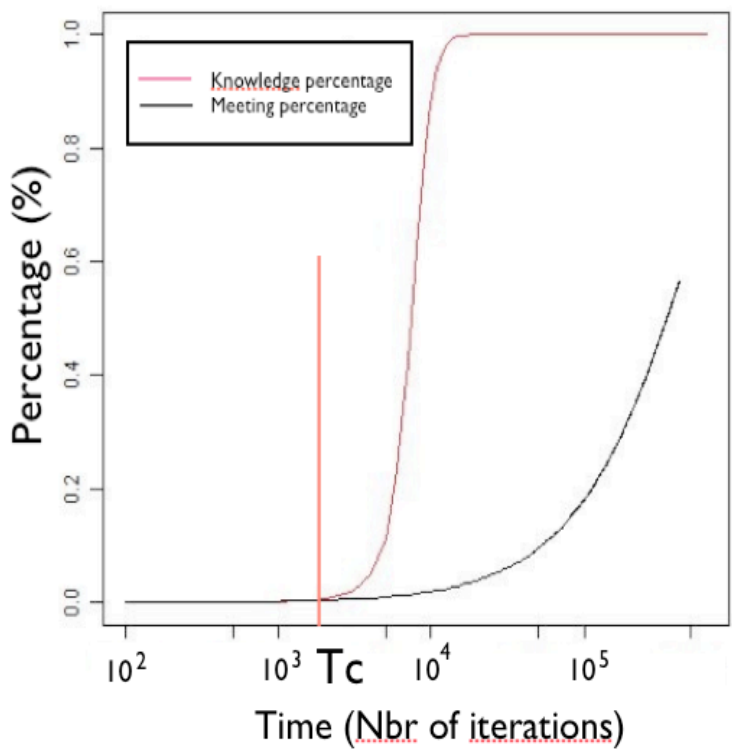

Figure 10: Evolution of Mutual Knowledge (red curve) and meeting rate (black curve) over time (logarithmic scale). The vertical red line shows the time of percolation (Tc). Percentage on the y-axis is the percentage of MK relation to the maximal MK where every agent knows everything about all of the other agents.

As we can see, below the percolation point Tc, the knowledge is not shared; agents may exchange some information but we do not have any propagation of this information. Above the threshold (Tc), the percentage of shared knowledge grows much more rapidly than the agents that meet. This means that even before all agents have met, every agent knows everything and this process has the properties of a percolation. This can be explained by the fact that when two agents meet, at maximum one edge is added to the meeting network (if the two agents have not met before). Conversely, up to 2(n-1) arcs can be added to the knowledge network since the two agents exchange their information.

The common points of the two networks are:

- The two curves have the same appearance (the knowledge level and the meeting proportion are close to zero at the start); only a few agents have met and information has not been propagated. From Tc (the percolation threshold) the curves rapidly increase and become stable and equal to 1 signifying that all the agents have met and all the information has been diffused throughout the population.

- This observation confirms our previous results: the propagation of information and the frequency of meeting are governed by a percolation process; the link percolation threshold is the same in the two networks, i.e. 0.001 for $10^3$ agents.

The differences are:
- The speed of information propagation is quicker than the meetings. The time necessary to reach full mutual knowledge is a lot shorter than the time for all the agents to meet. As mentioned previously, when two agents meet, a maximum of one edge is added to the meeting network (if the two agents have not met before). Conversely, up to 2(n-1) arcs can be added to the knowledge network since the two agents exchange their information.
- The time necessary to reach the percolation threshold is shorter in the meeting network, but we cannot currently explain this difference, however the time taken for the two networks to be totally connected is negligible.
- The link percolation threshold is the same in the two networks (knowledge and meeting networks), i.e. 0.001 for 1000 agents.

In the following section we analyse how percolation is sensitive to network parameters.

**5.4 Communities structure**

In order to detect the structure of communities in both studied series of networks (the meeting and the knowledge network) (size $10^3$) we used Pajek visualization. Figure 11 shows the result before, during and after the percolation threshold.

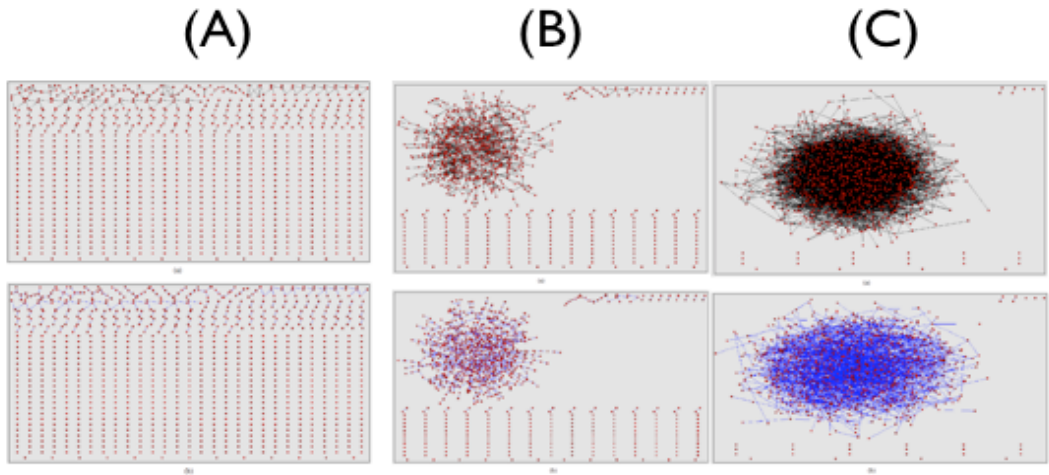

Figure 11: The knowledge network (up) / meeting network (down) (A) at the beginning of the simulation, (B) at the percolation time Tc and (C) after the percolation time. ($N = 10^3$).

For each of the networks, at the beginning of the simulation, the network is composed of small clusters (of which the majority are isolated nodes). When we reach the percolation threshold a large cluster appears, and in advancing the simulation all small clusters disappear and the large cluster envelops the entire network. These results show the presence of a percolation phenomenon.

**5.5 Effect of the number of agents**

In order to study the effect of the number of agents (N), we performed simulations with $N = 10^3$, $10^4$ and $10^5$, d = 5000, m = 1, overhearing = 0 and forgetting = 0.
Following table 2, we note that for the knowledge, results show that the value for the percolation threshold $p_c$ vary with $1/N$. Indeed it is exactly the probability at which the phase transition leads to the emergence of a large cluster appearing in random graphs, as shown by Erdos and Renyi (1959).

The percolation threshold (link or node) is reached more rapidly when N increases.

| N | $10^3$ | | | $10^4$ | | | $10^5$ | | |
|---|---|---|---|---|---|---|---|---|---|
| | Pbc | Tc | Nb | Pc | Tc | Nb | Pc | Tc | Nb |
| | 0,001 | 18000 | 999 | 0,0001 | 1800 | 9999 | 0,00001 | 180 | 99999 |

Table 2 : Percolation threshold values (Nb) and the proportion of active arcs (Pbc) and time (Tc) in knowledge networks of sizes $10^3$, $10^4$ and $10^5$.

The emergence time of a large cluster ($t_c$) is shorter when N is larger and we reach the percolation threshold more quickly when N increases.

**5.6 Effect of density**

We note that the denser the population, the higher the meeting frequency (figure 12 B) and thus information propagation is more rapid (figure 12 A).

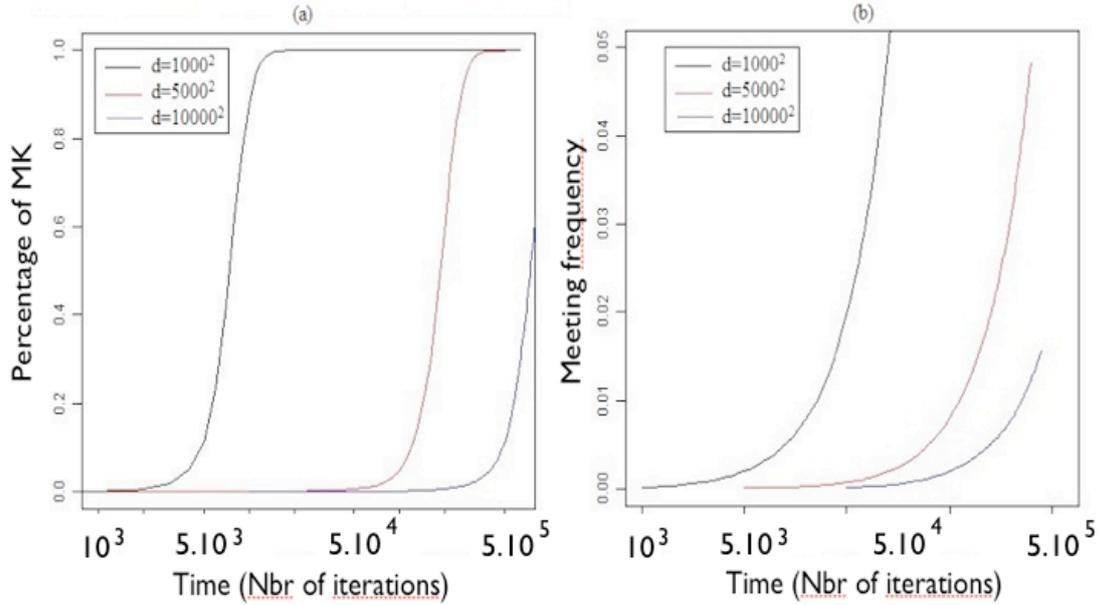

Figure 12: Information propagation (A) and meeting frequency (B) for d = $1000^2, 5000^2$ and $10000^2$. Percentage = percentage of MK in relation to the maximal MK.

We can see that density does not influence the percolation threshold value in the meeting network (node and edge percolation) or in the knowledge network (node and arc percolation). However it does influence the time needed to reach the threshold; this being shorter for a higher density environment (Table 3).

|   |    | $1000^2$ |     |       | $5000^2$ |       |       | $10000^2$ |       |       |
|---|----|----------|-----|-------|----------|-------|-------|-----------|-------|-------|
|   |    | Pbc      | Tc  | Psc   | Pbc      | Tc    | Psc   | Pbc       | Tc    | Psc   |
| d | RC | 0,001    | 800 | —     | 0,001    | 18000 | —     | 0,001     | 70000 | —     |
|   | RR | 0,001    | 500 | 0,636 | 0,001    | 15000 | 0,686 | 0,001     | 60000 | 0,71  |

Table 3: Percolation threshold values of the proportion of links (Pbc) and active nodes (Psc) and time (Tc) in knowledge networks (RC) and in meeting networks (RR) for different values of d (d=$1000^2$, d=$5000^2$ and d=$10000^2$)

**5.7 Effect of overhearing**
In order to observe this effect we ran a series of simulations, each time changing the extent of the overhearing (overhearing = 0, 1, 2, 3, 4) whilst keeping the other variables stable (N = $10^3$, d = $5000^2$, m = 1, forgetting = 0).
Figure 13 shows the effect of agents overhearing on the emergence of mutual knowledge. We note that when overhearing is higher, the propagation of information is faster and the agents produce mutual knowledge more quickly.

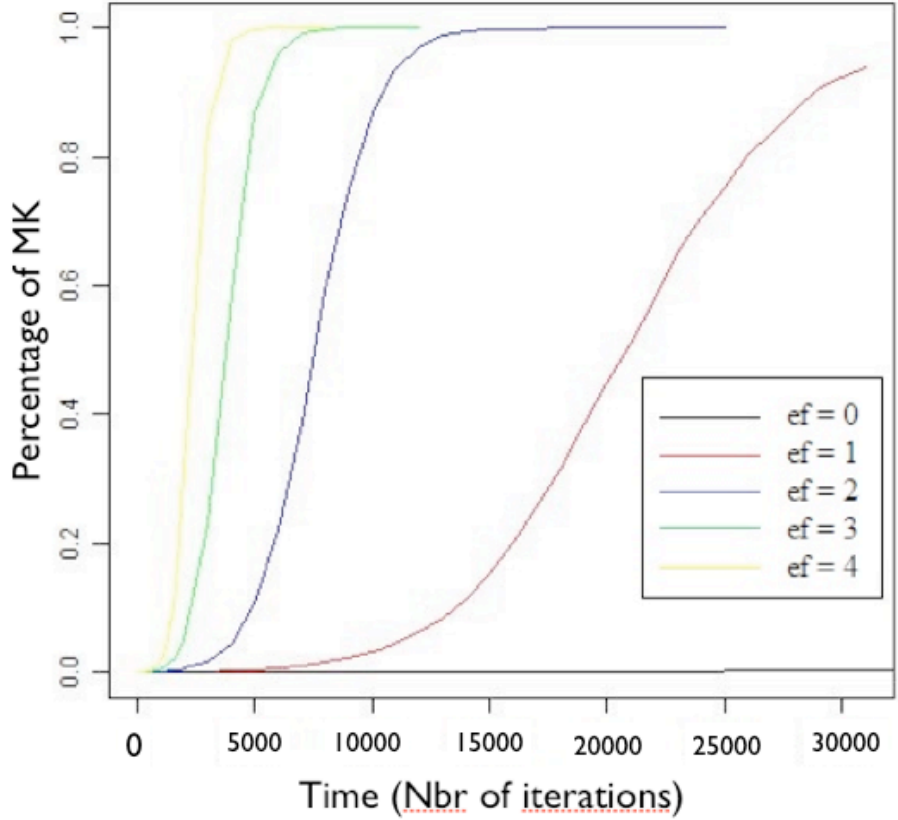

Figure 13: Evolution of mutual knowledge as a function of overhearing. Percentage of MK = percentage of MK in relation to the maximal MK (ef = overhearing distance).

|    |    | Pbc | Tc | Psc | Pbc | Tc | Psc | Pbc | Tc | Psc |
|----|----|-----|-----|-----|-----|-----|-----|-----|-----|-----|
| ef | RC | \multicolumn{3}{c} 0 | | | 1 | | | 2 | | |
|    |    | 0,001 | 18000 | — | 0,001 | 2000 | — | 0,001 | 800 | — |
|    |    | 3 | | | 4 | | | | | |
|    |    | 0,001 | 400 | — | 0,001 | 300 | — | | | |
|    | RR | 0 | | | 1 | | | 2 | | |
|    |    | 0,001 | 15000 | 0,686 | 0,001 | 1500 | 0,651 | 0,001 | 600 | 0,689 |
|    |    | 3 | | | 4 | | | | | |
|    |    | 0,001 | 300 | 0,644 | 0,001 | 200 | 0,73 | | | |

Table 4: Percolation threshold values of the proportion of links (Pbc), active nodes (Psc) and time (Tc) in knowledge networks (RC) and meeting networks (RR) for different values of overhearing (overhearing = 0, 1, 2, 3 and 4).

In both networks the percolation threshold is reached more rapidly when overhearing is higher, but the threshold value is always the same.
Concerning the structural properties of the networks, overhearing has no influence on the shape of the representative curves. However, there is a difference in the number of iterations that it takes to produce the curve, this shows that mutual knowledge emerges more quickly when overhearing is higher.
In the two following sections, we try to assess the robustness properties of mutual knowledge processes. We have already seen that MK follows a percolation process with universal percolation singularities, as characterized by very fast information propagation and threshold values, etc. but we found worthwhile to investigate how such process is robust to destructive processes such as probability to forget, or unexpected information modifications.

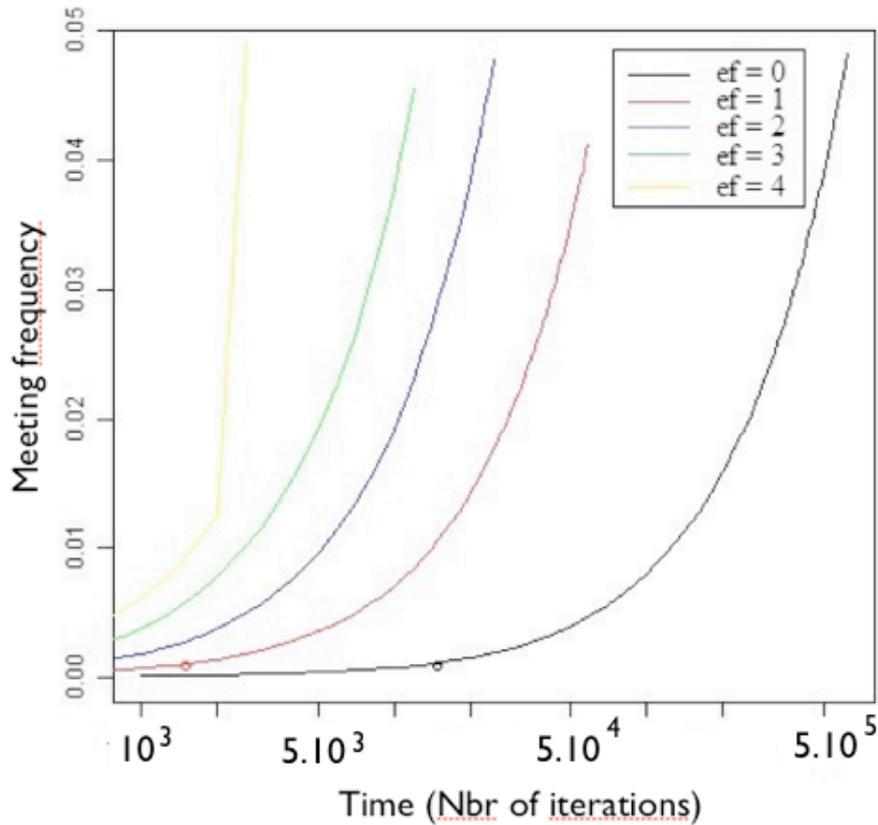

Figure 14: Meeting frequency as a function of overhearing distance (ef) and time.

**5.8 Effect of the probability to forget on information propagation**

Studying the effect of forgetting allows us to analyse network robustness against the random destruction of some of its links (which is a way to topologically simulate the probability to forget). The study is restricted to the knowledge network since forgetting concerns knowledge and not meeting. Figure 15 shows that as the probability to forget rises, the level of knowledge is weaker and the agents no longer produce mutual knowledge.

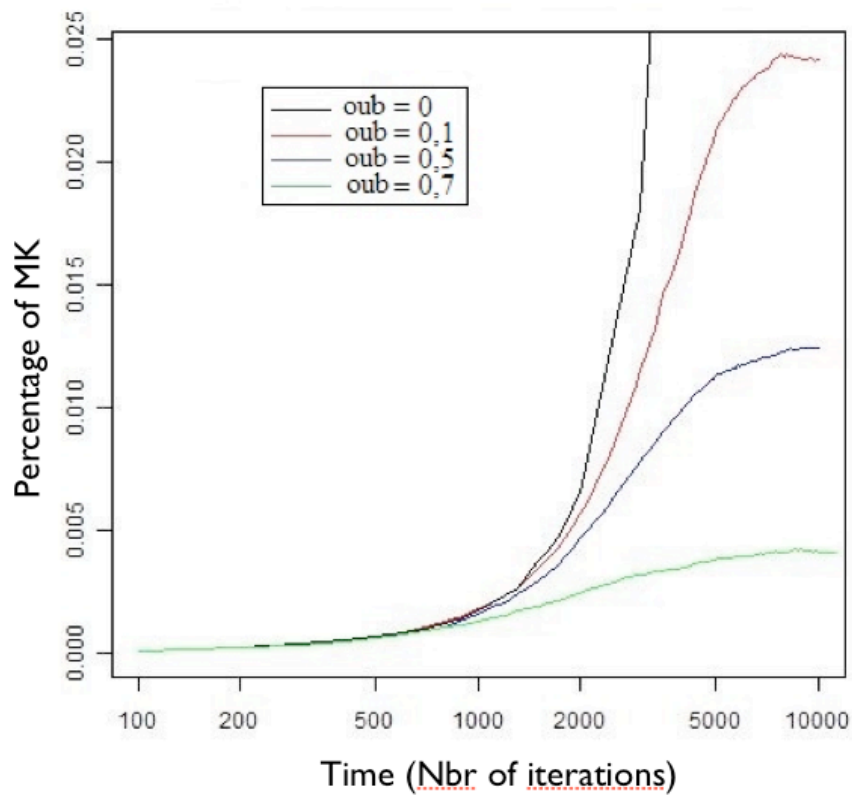

Figure 15: Evolution of mutual knowledge as a function of the forgetting probability (oub). Overhearing= 3, Density= 5000x5000, Size of population= 1000.

To better assess this result we calculated the knowledge level obtained in the same time interval according to the probability of forgetting (figure 16). The graph shows a rapid decrease as the probability of forgetting increases.

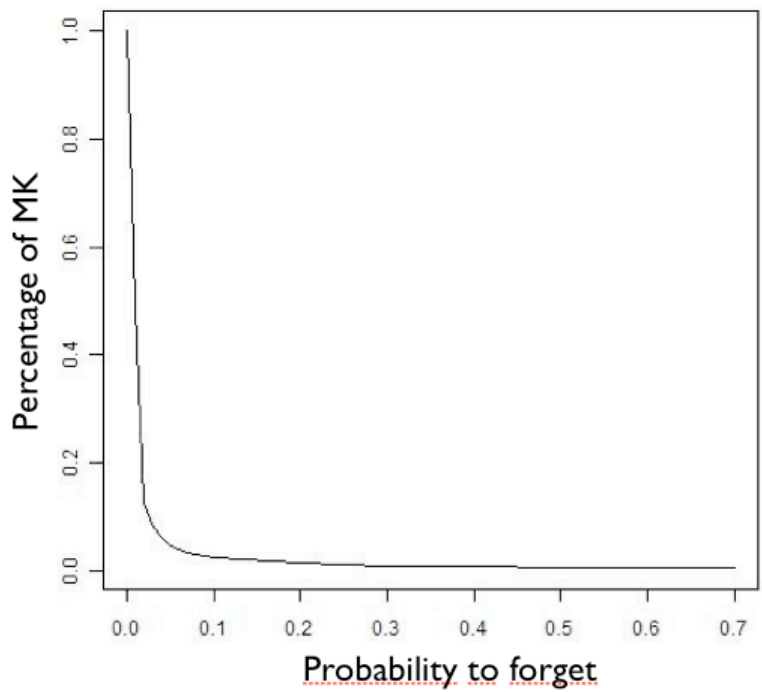

Figure 16 - Incidence of forgetting (Probability to forget) on the emergence of mutual knowledge (MK). Overhearing= 3, Density= 5000x5000, Size of population= 1000.
These results allow us to conclude that the network is weakly resistant to the removal of some arcs (low forgetting probability). It does not support a frequent removal of arcs and becomes disconnected; the information remains local and does not propagate. The robustness of the network is therefore not very strong.

**5.9 Effect of changing information**

Agents' group memberships are updated during the simulation. We have measured the gap between the knowledge of agents and that of the 'real-world' ('real-world' in this case is the list of true information at each instant. The agents' knowledge is the information held by agents that may disagree with the real world).
We first studied the effect of the frequency of changing information. As intuitively expected, the gap closes when the frequency of changing information is lower (figure 17).

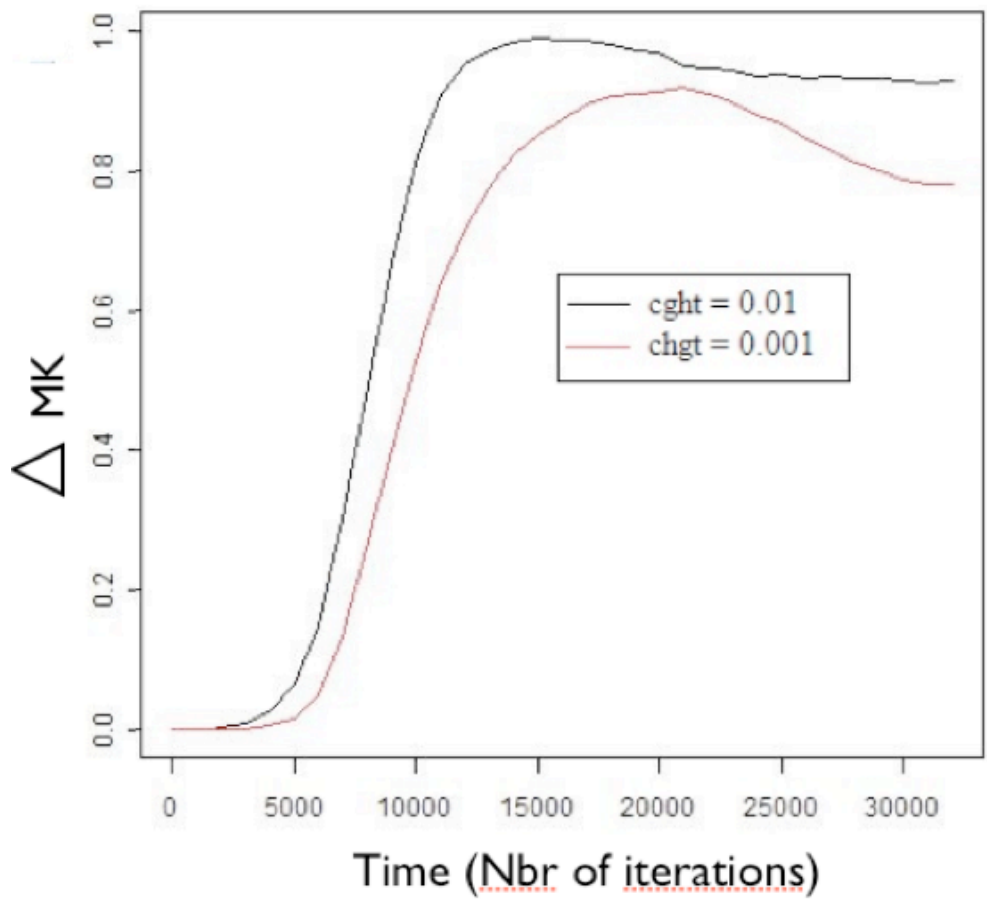

Figure 17: Influence of the frequency of changing information (cght) on the perception of the environment. Delta MK = the gap between agent MK and real MK. Overhearing= 3, Density= 5000x5000, Size of population= 1000.
We can see that even a very weak probability of changing information induces a permanent gap between the real knowledge held by agents and the mutual knowledge. The process is therefore very fragile to loss of agent memory.
We then investigate if this process of loosing memory depends on the size of the population. From figure 18, we can see that increasing the size of the population increases the gap between agents' knowledge and that of the 'real-world'. Thus, the higher number of individuals that an agent knows, the harder it is to update incorrect information.

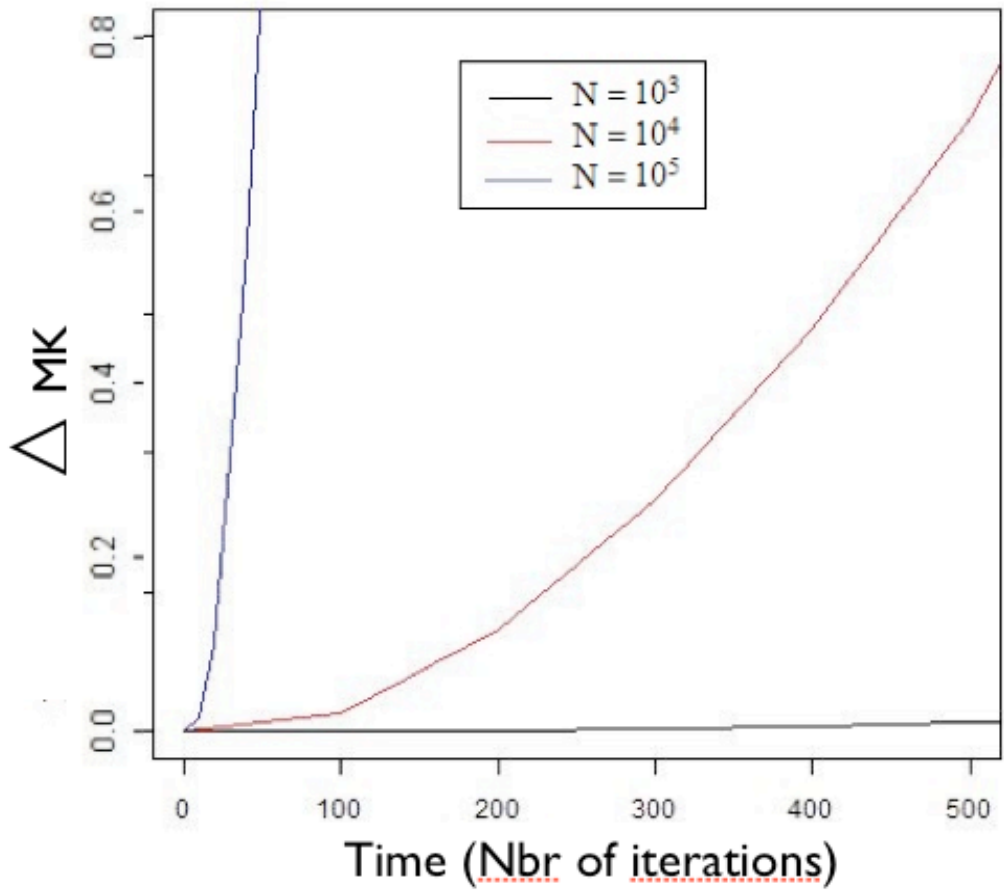

Figure 18 - Influence of population size (N) on on the gap between the knowledge of agents and that of the 'real-world' (Delta MK). Overhearing= 3, Density= 5000x5000, Size of population= 1000.

With a low population size (N=1000), the loss of individual memory has almost no effect on the quality of mutual knowledge; agents update their knowledge more rapidly than they forget their knowledge.

Finally we analyse how the population density affects mutual knowledge. Figure 19 demonstrates the importance of the environment where knowledge evolves over time. In situations where information must be regularly updated, the simulation shows that one of the most influential variables is population density; the denser the population, the faster the propagation of false information.

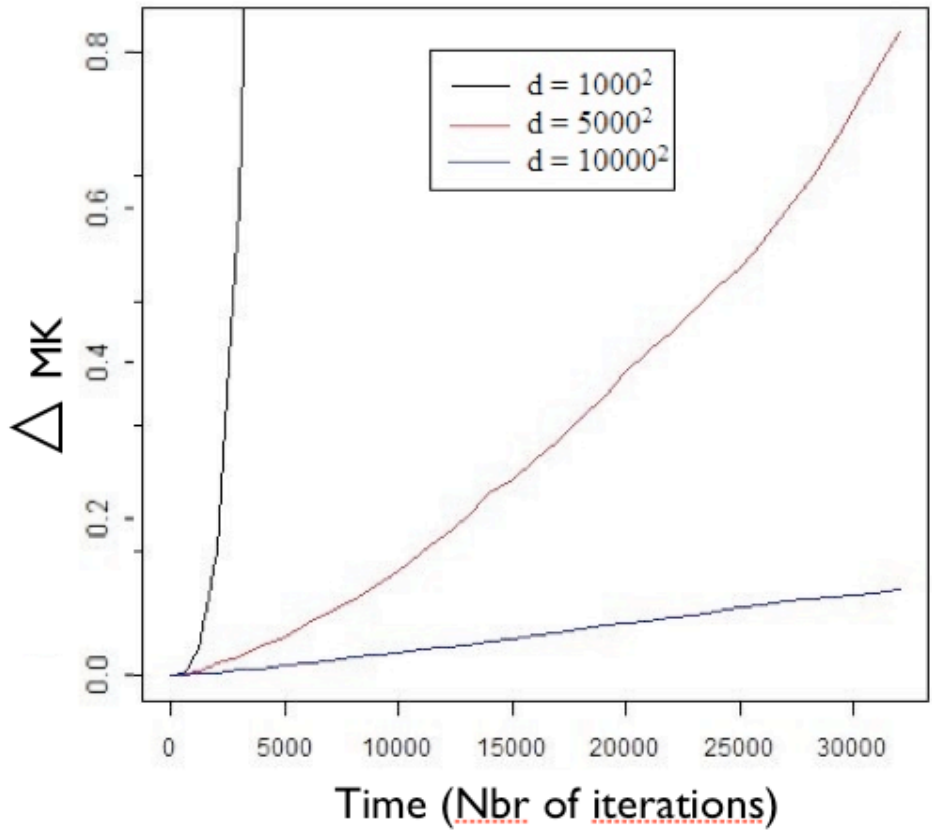

Figure 19: Gap between agent's mutual knowledge and the real-world mutual knowledge (Delta MK) for different environmental dimensions. Overhearing= 3, Size of population= 1000)

Contrary to what was expected, overhearing increases the gap between agents' knowledge and the real-world (figure 20). However, this gap starts to decrease when the network becomes fully connected, i.e. when all agents are known. So instead of improving the capacity of agents to update and correct false information, overhearing favours the spread of false information, but after a certain time interval, which is shorter when overhearing is higher. This effect will be reversed by playing an important role in updating false knowledge.

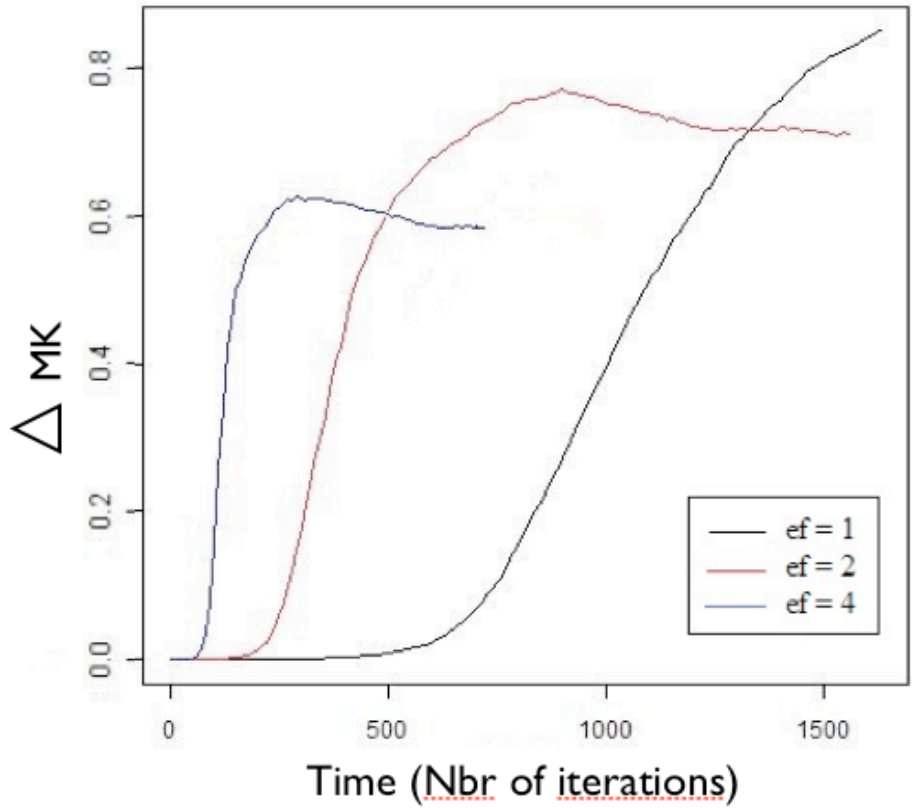

Figure 20: Gap between mutual knowledge and that of the real world (Delta MK). A small overhearing abilities (ef= 1) affects the mutual knowledge process more than an larger overhearing ability (ef=4).

The study of robustness properties of social networks is very important because, as we frequently see in real systems, security issues of complex socio-technical networks are often related to the emergence of mutual knowledge.

These results show that the robustness of the social network depends on several factors, such as the size of the population, radius of overhearing, and the density of the population.

We investigate this process through two mechanisms: the probability of forgetting agent knowledge and modifying agents' internal knowledge.

We have seen that:
- The network is robust to forgetting as soon as the probability of forgetting does not exceed 0.1
- Randomly changing agents' internal information, if the probability of change is low (<0.01) and the population density is small, does not impact mutual knowledge
- Contrarily to expectations, a low overhearing ability (ef= 1) has a more important effect than a high overhearing ability (ef-4), on mutual knowledge. This result was interpreted by assuming that a higher overhearing ability more greatly affects the propagation of false information than repairing it through agent's meeting.

Following these results, it appears that social networks that rely on MK to share their knowledge are robust but only if false information is marginal. We intend by deleter information, the processes of forgetting or random agent information transformation.

If not, the process of sharing information through meeting is more favourable to the propagation of false information than to the stabilisation of a true MK.

**5.10 Summary of results**

The meeting networks show a mixed (nodes and link) percolation phenomenon, whereas the knowledge network only shows the arc percolation phenomenon.
At the start of their evolution, networks are composed of small isolated clusters and at the percolation threshold a large cluster appears that eventually envelops the entire network.
The value for the link percolation threshold is the same in the two networks and does not depend on the size of the network.
The probability at the percolation threshold for link percolation is inversely proportional to the size of the network; this result has also been demonstrated by Erdos and Renyi for random graphs.
In both networks, the degree distribution follows a power law; this is one of the principal characteristics of scale-free networks.

**6 Conclusions**

The aim of this paper was to question the theoretical nature of the emergence of MK. Emergent behaviour usually refers to spontaneous outcome due to the interaction between many actors in critical situations. Such emergent behaviours allow people to efficiently cooperate in complex socio technical systems such as Air Traffic Control and regulation centres, etc. Multi-agent simulations have been extensively used both to reproduce such emergent behaviours and as tools to design robust social networks.

Nevertheless, identifying MK as a percolation process has never been strictly proved. Percolation can be seen as a specific and interesting type of emergence because it exhibits very specific properties that are independent of the network's characteristics. As examples, we have seen that at the percolation threshold, the size of the connected network (also called giant network) grows as a power law; that giant connected networks show fractal dimensions, and that the robustness of percolated networks can be assessed.
From a social theory point of view, considering percolation as a specific cooperative property in social networks is very interesting since it provides a structural framework of the emergence of global properties that go beyond individual representations. Thus, it gives us a theoretical framework to understand emergence in real social networks. We therefore investigate how percolation can also be seen as a mark of optimization. Optimization is often considered as a driving force for the evolution of biological and social structures. We can therefore consider that the emergence of MK through communication processes is an evolution towards very efficient collective structures.

Our comparison was based on the work of Barratt and his colleagues [2] where they propose criteria for the appearance of a giant cluster in graphs that have a local tree structure with no cycles. Results clearly show that the dynamics of the emergence of MK conform to the critical percolation condition. Conditions on heterogeneity parameters are respected and non-linear behaviour is characteristic of emergent systems.

Concerning the robustness, a network's robustness was tested with the random destruction of some of its arcs (simulating the process of forgetting) as well as with changing node information (simulating the process of cheating). Results confirm what has often been observed in real situations; that the emergence of robust MK is more easily obtained with networks of small population with no deterrent phenomena such as forgetting or false information propagation. The emergence of MK appears to be very sensitive to deterrent processes. If we view this in light of communication theory, we could say that the emergence of MK (or efficient cooperation) may occur very rapidly in a cooperative network, but this MK can drop quickly if actors do not follow the Grice's maxims of good cooperation (specifically the sincerity condition) [13]. These results are coherent with our empirical field observations that showed that the efficiency of team cooperation drastically falls as soon as overhearing is reduced (e.g. due to a noisy working environment) or as soon as actors hide information [23].